\documentclass[12pt]{iopart}

\usepackage{cite}
\usepackage{iopams}
\usepackage{epsfig}
\usepackage{subfigure}

\begin{document}

\title[Metamaterials with dual-band magnetic resonances]{Composite metamaterials with dual-band magnetic resonances in the terahertz frequency regime}

\author{Ming Li, Zhenchao Wen, Jinxin Fu, Xu Fang, Yaomin Dai, Rongjuan Liu, Xiufeng Han, and Xianggang Qiu\footnote[1]{Electronic mail: xgqiu@aphy.iphy.ac.cn}}

\address{Institute of Physics and Beijing National
Laboratory for Condensed Matter Physics, Chinese Academy of
Sciences, Beijing 100190, China}

\date{\today}

\begin{abstract}
Composite metamaterials(CMMs) combining a subwavelength metallic
hole array (i.e. one-layer fishnet structure) and an array of
split-ring resonators(SRRs) on the same board are fabricated with
gold films on silicon wafer. Transmission measurements of the CMMs
in the terahertz range have been performed. Dual-band magnetic
resonances, namely, a LC resonance at 4.40 THz and an additional
magnetic resonance at 8.64 THz originating from the antiparallel
current in wire pairs in the CMMs are observed when the electrical
field polarization of the incident light is parallel to the gap of
the component SRR. The numerical simulations agree well with the
experimental results and further clarify the nature of the dual-band
magnetic resonances.
\end{abstract}


\section{Introduction}
Over the decades, the desire to manipulate the state of the
electromagnetic wave propagation has been attractive to researchers
in a variety of fields such as optics and
chemistry~\cite{science01,halas}. Metamaterials, the artificial
materials with subwavelength structures, could exhibit some unusual
properties which are critical for a number of potential applications
such as negative refraction, superlens and electromagnetic
cloak~\cite{apl,superlens,cloak}. The first negative index
metamaterials(NIMs) were demonstrated in the microwave frequency by
Smith et al.~\cite{smith}, who designed a composite structure
consisting of Pendry's metal wires and double split-ring
resonators(SRRs)~\cite{wire,srr}. Subsequently, negative effective
permeability has been achieved at terahertz and infrared frequencies
via SRRs with reduced sizes~\cite{science,prl,german}. Meanwhile, to
hurdle the large metal losses at higher frequencies and the
difficulty of experimental fabrication~\cite{zongsu}, alternative
designs are introduced, showing that pairs of metal wires or metal
plates, as well as multilayer fishnet structures can provide
magnetic resonance at infrared and even visible
frequencies~\cite{shalaev,zhangs, zhangx,visible}. All the NIMs
mentioned above exhibit the negative index behavior only at a single
band. It has been proposed by D.H. Kwon et al. that dual-band
negative index property can be realized in composite metamaterials
(CMMs) with two periodic substructures~\cite{dual}.They reported the
first numerical validation of dual-band NIM performance at two
separated frequency bands under the same polarization in the
near-infrared region. Also, C. Yan et al. investigated two
electromagnetic resonance phenomena in the microwave regime with
numerical simulations~\cite{yan}. In this paper, by embedding the
SRRs within the dielectric holes (which are air in our experiment)
of the fishnet structure, we obtain dual-band magnetic resonances in
the terahertz range for a linearly polarized light. Our results may
be extended to other frequencies and have some potential
applications in designing novel metamaterials.

\section{Sample preparation} \label{MT}
Shown in Fig.1(c),(d) are the schematic view of one unit cell and
the the scanning electron microscope (SEM) image of the designed CMM
(sample C) comprising SRRs embedded in holes of a perforated Au
film. For comparison, a metal film perforated with a periodic array
of subwavelength square holes(sample A, schematic in Fig.1(a)), and
a quadratic array of SRRs (sample B, schematic in Fig.1(b))are also
prepared. The lattice constants ($\emph{l}$) of Sample A, B, C are
all 14 $\mu$m. The side length of a square hole for the fishnet is
11 $\mu$m and the length of an arm (d) in SRR is 7 $\mu$m. The
dimensions of the square holes and SRRs embedded in them for sample
C are the same as those of sample A and sample B.

\begin {figure}
\begin{center}
\includegraphics[width=3.5in]{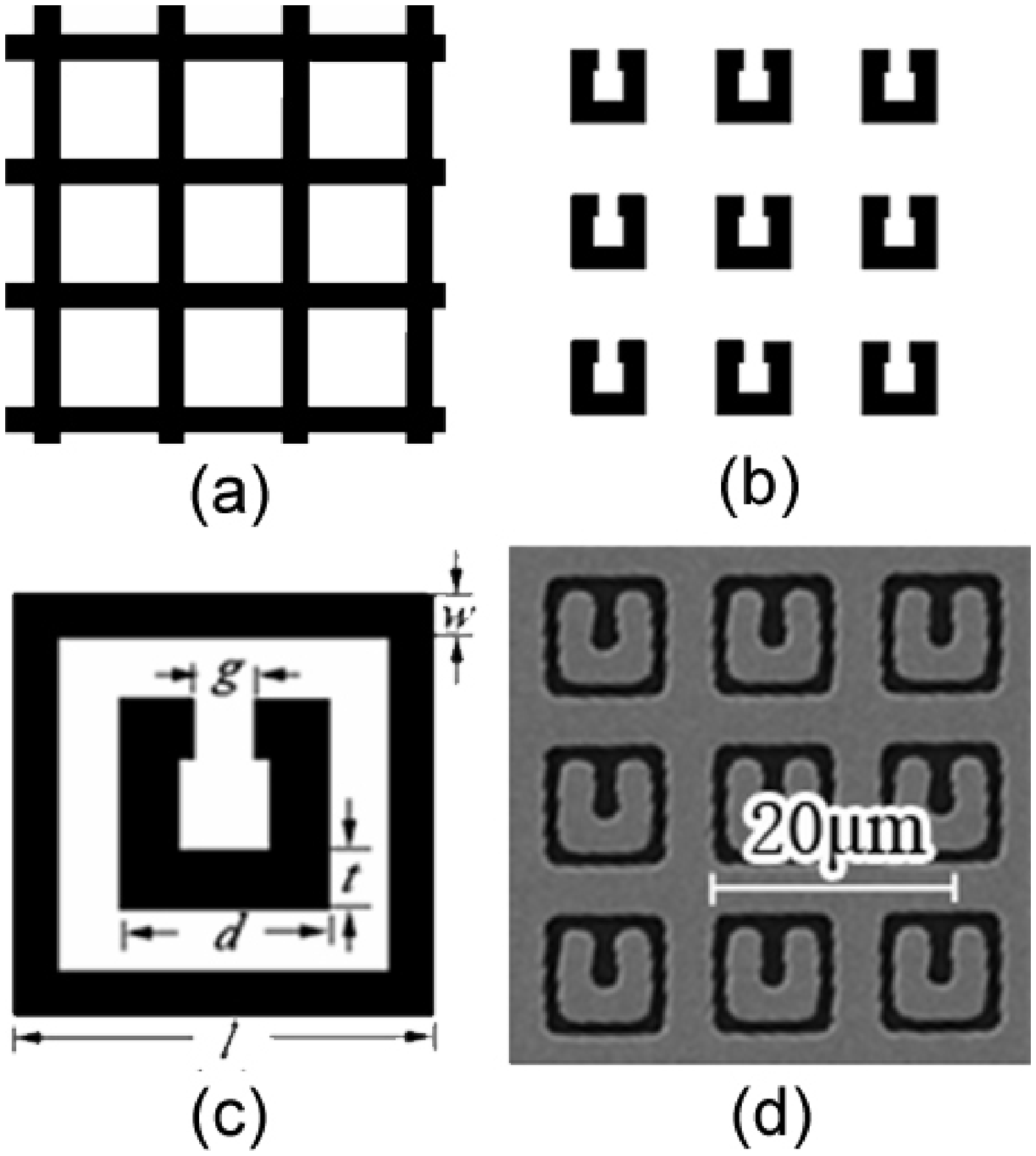}
\caption{(a)The schematic of Sample A. (b)The schematic of Sample B.
(c)The schematic of one element of Sample C with parameters
$\emph{l}$=14 $\mu$m, d=7 $\mu$m, g=2 $\mu$m, and t=w=2 $\mu$m. (d)
The SEM picture of sample C.}
\end{center}
\end{figure}

These samples were fabricated with gold films of 300nm thick
deposited on silicon substrates and patterned with ultraviolet
photography. To improve the adhesion between the gold film and the
silicon substrate, a thin layer of 5 nm tantalum(Ta) was first
deposited on the surface of the Si wafer. Each array used for the
experiment here has an area of 7$\times$7 mm$^{2}$.

\section{Spectroscopic measurements and numerical analysis}
We acquired the normal-incidence transmittance spectra of the
samples from 2 to 12 THz with an ABB Bomem DA8 Fourier transform
infrared spectrometer. Horizontal and vertical polarizations of
light were used for the measurements. In horizontal polarization the
electric field of the incident light is parallel to the featuring
gap of the SRR, while in vertical polarization it is perpendicular
to the SRR gap.

The normal-incidence transmission spectra of sample A, B, and C with
horizontal and vertical polarization are shown in Fig.2 (a) and (c)
respectively. Using a commercial software CST Microwave Studio we
numerically investigated the transmission of one unit cell of these
arrays. Considering the overall shape and the positions of peaks and
dips, the simulations are in good agreement with the experimental
data, as shown in Fig. 2(b) and (d).

For sample A, the peaks at frequencies of 6.17 THz and 8.85 THz are
attributed to the enhanced transmission assisted by surface plasmon
polaritons of Si(1,0) and Si(1,1) at the metal and silicon
interfaces~\cite{fang}. These experimental results agree excellently
with what the momentum matching condition indicates~\cite{ebbeson}:
$\textbf{k}$$_{sp}$=$\textbf{k}$$_{0}$ $\pm$ i$\textbf{G}$$_{x}$
$\pm$ j$\textbf{G}$$_{y}$, where $\textbf{k}$$_{sp}$ is the wave
vector of SPPs, $\textbf{k}$$_{0}$ is the component of the incident
wave vector in the metallic film plane, and $\textbf{G}$$_{x}$ and
$\textbf{G}$$_{y}$ are the reciprocal vectors of the fishnet array.
The integers, i and j, are used to designate the resonant peaks.

The dip at 4.40 THz at the curve of sample B in Fig.2 (a) suggests
the LC resonance that arises from a circulating current in the coil
induced by the external horizontally polarized electric field. This
magnetic resonance vanishes for the vertical polarization, leaving
behind only the Mie resonance of the SRR around 6.51 THz frequency
in Fig.2 (c). Also, the weak dip at 8.92 THz in Fig.2 (a) originates
from Mie resonance, which is due to the depolarization field of the
short axis, i.e., the width of the SRR arms~\cite{prl}.

\begin {figure}
\begin{center}
\includegraphics[width=4.8in]{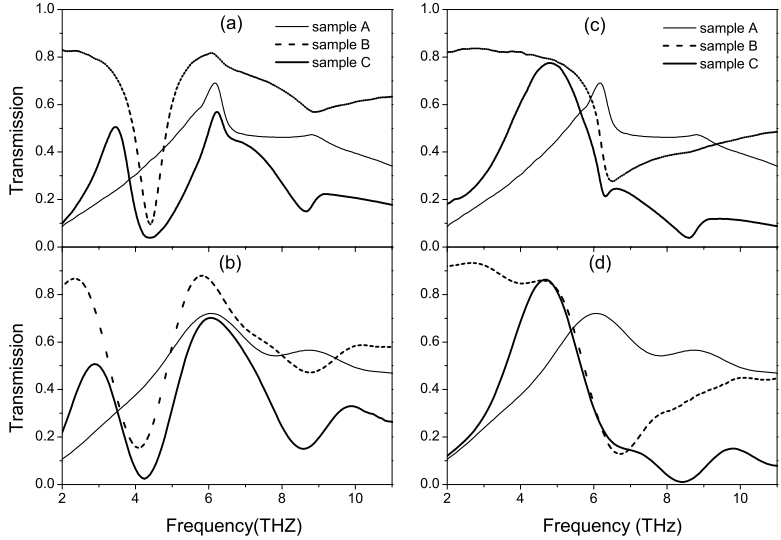}
\caption{Transmission spectra of sample A, B, and C. (a) and (c) are
the experiment results of horizontal and vertical polarization
respectively, while (b) and (d)are the corresponding results of
numerical calculation for each polarization. For vertical
polarization, E-field is perpendicular to the SRR gap.}
\end{center}
\end{figure}

The spectra of sample C obviously inherit some characteristics of
sample A and B. The reduced average transmission of Sample C in the
low frequency is expected as a result of the existence of the
forbidden transmission band in the long wavelength region of sample
A. Besides, The location of the LC resonance for the horizontal
polarization does not change, compared to that of sample B. The
calculated field distribution (not shown here) for horizontal
polarization suggests that the electronic excitation still localizes
at the SRR's featuring gap. The dip at 8.64 THz of sample C,
however, is attributed to the magnetic resonance excitation induced
by the horizontally polarized electric field, though its position is
close to the Mie resonance of SRRs in sample B at 8.92 THz as shown
in Fig.2(a). This can be verified in the calculated current density
distribution in Fig.3 (a). The charge accumulation is opposite at
the corresponding ends of the SRR bottom edge and the nearby wire of
fishnet. Therefore, the antiparallel electric current in the two
wires could excite magnetic field normal to the CMM plane. It is
also interesting to find that the resonant wavelength (34.7 $\mu$m)
is approximately twice the effective optical length of the
horizontal SRR arm: $\lambda$$\approx$2d*n$_{si}$, where
d*$\approx$5 $\mu$m and n$_{si}$=3.42 is the refractive index of the
silicon substrate~\cite{france}.

A further study strengthens the above interpretation. It is noticed
that at vertical polarization a dip emerges at 8.59 THz for sample
C, which vanishes in sample B where there are SRRs alone, as shown
in Fig.2(c). We find that its position is almost the same as the
8.64 THz dip for sample C at horizontal polarization. Actually, they
both have the same physical origin. The 8.59 THz dip for sample C at
vertical E-field polarization arises from the magnetic resonance of
parallel metallic wires with antiparallel electric current. In this
configuration, the wire pairs are the SRR arms and their neighboring
fishnet wires. The induced current density distribution shown in
Fig. 3(b) definitely characterizes such features: the charge
accumulation oscillates in anti-phase at the SRR arms and the
neighboring fishnet wires, and thus induces the antiparallel current
which results in a magnetic field normal to the CMM surface.

\begin {figure}[!]
\begin{center}
\includegraphics[width=3.5in]{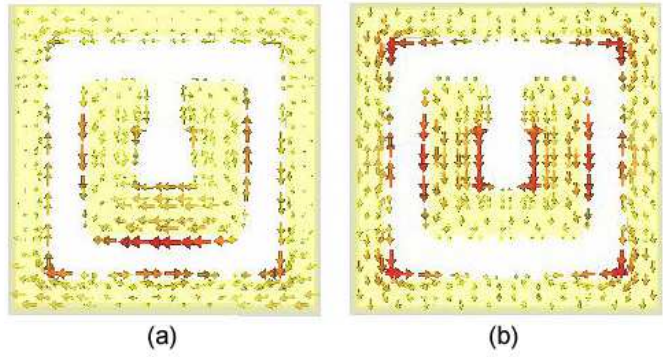}
\caption{(Color online) Induced current density distribution on the
interface between the metal film and Si substrate in sample C. (a)
represents current density distribution at 8.64 THz for horizontal
polarization. (b) shows current density distribution at 8.59 THz for
vertical polarization. Both magnetic resonances at 8.64 THz and 8.59
THz originate from antiparallel currents induced by external E-field
in the corresponding wire pairs.}
\end{center}
\end{figure}

To further confirm the above findings, we fabricated sample D, which
has the same building blocks as sample C but with different
dimensions. The SRRs in sample D remain the same size as that in
sample C, but the lattice constant is enlarged to 22 $\mu$m, and the
side length of holes becomes 17 $\mu$m. For comparison, we put the
measured spectra of samples C and D together as shown in Fig.4,
where (a) and (b) present the corresponding normal-incidence
transmission spectra for horizontal and vertical polarization
respectively. In both pictures, the thick solid lines are the
experimental results of sample D, and the thin solid lines represent
spectra of sample C. It is shown that the spectra of samples C and D
have similar configuration. With numerical simulations, we could
verify the nature of the dips in the spectra of sample D, which have
the same origin as those in sample C. The slight red shift of the
spectrum of sample D with respect to that of sample C probably
results from the weaker coupling between the SRRs and the
neighboring wires of the fishnet in sample D, since the distances
between SRRs and the neighboring wires of the fishnet structure are
larger in sample D than those in sample C.

\begin {figure}
\begin{center}
\includegraphics[width=6.0in]{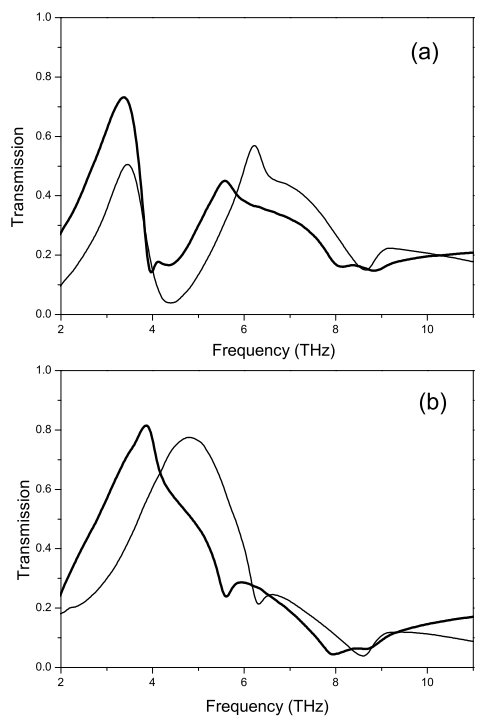}
\caption{Transmission spectra of sample D, compared with sample C.
(a) transmission under the horizontal electrical field polarization;
(b) transmission under the vertical electrical field polarization.
In both pictures, the thick solid lines are the experimental results
of sample D, and the thin solid lines represent spectra of sample
C.}
\end{center}
\end{figure}

\section{Conclusions}   \label{SUM}
In summary, composite metamaterials with SRRs embedded within the
holes of one-layer fishnet structure are fabricated. Experimental
and numerical studies indicate that there are dual-band magnetic
resonaces induced by external electric field in the CMMs for the
normal-incidence light. Dual-band negative permeability could occur
for such CMMs when the incident waves propagate parallel to the
sample surface. However, the large dielectric constant of silicon
makes it difficult to extract the accurate electromagnetic
parameters in the terahertz region~\cite{prb}. Further research
about this topic is expected to bring about potential application in
in designing novel negative index metamaterials and thus in the
engineering of controlling the light propagation properties.

\section{Acknowledgement}
The authors thank Z.Y. Li for beneficial advice and discussion, and
J.J. Li for technical help. This work is supported by National
Science Foundation of China (No.10674168), the MOST of China(973
project No. 2006CB601006).

\end{document}